\documentclass[conference]{IEEEtran}
\IEEEoverridecommandlockouts

\usepackage{cite}
\usepackage{amsmath,amssymb,amsfonts}
\usepackage{algorithmic}
\usepackage{graphicx}
\usepackage{textcomp}
\usepackage{xcolor}
\usepackage{ifthen}

\def\BibTeX{{\rm B\kern-.05em{\sc i\kern-.025em b}\kern-.08em
T\kern-.1667em\lower.7ex\hbox{E}\kern-.125emX}}

\usepackage[colorlinks=true,
citecolor=blue,    % 正文文献引用[1]颜色
linkcolor=blue,    % 图、表、章节交叉引用颜色
urlcolor=blue]{hyperref}

\begin{document}

\title{Token Importance-Aware Semantic Communication Enabled Robust Video Transmission}

\author{
\IEEEauthorblockN{
Xiwen Zhang$^{\ddagger}$,
Yubo Peng$^{*\dagger\ddagger}$,
Luping Xiang$^{*\dagger\ddagger}$,
and Kun Yang$^{*\dagger\ddagger}$
}
\IEEEauthorblockA{
$^{*}$State Key Laboratory of Novel Software Technology, Nanjing University, Nanjing 210008, China\\
$^{\dagger}$Institute of Intelligent Networks and Communications (NINE), Nanjing University (Suzhou Campus), Suzhou 215163, China\\
$^{\ddagger}$School of Intelligent Software and Engineering, Nanjing University (Suzhou Campus), Suzhou 215163, China\\
Email: 241880592@smail.nju.edu.cn, ybpeng@smail.nju.edu.cn, luping.xiang@nju.edu.cn, kunyang@nju.edu.cn\\
\textit{Corresponding author: Luping Xiang}
}
}

\maketitle

\begin{abstract}

Video applications are increasingly prevalent, yet their high data volume and temporal dependency make robust wireless transmission challenging under limited communication resources. To address this issue, we propose a token importance-aware semantic communication (Ti-SemCom) framework for bandwidth-efficient video transmission. Specifically, object and motion information is jointly extracted to identify communication-relevant regions, which are hierarchically associated with semantic tokens according to their importance. The transmitted tokens are then dynamically selected based on both token importance and the available bandwidth, prioritizing more informative video content under resource constraints. At the receiver, untransmitted regions reuse recently reconstructed references to further reduce redundant transmission. Experiments on the CAVIAR dataset show that Ti-SemCom can achieve an 88.3\% reduction in communication cost while maintaining competitive reconstruction quality.

\end{abstract}


\section{Introduction}

With the growing use of video applications, large volumes of visual data need to be continuously transmitted over wireless networks. The high data rate of video places considerable demands on these networks, especially under limited bandwidth and adverse channel conditions. When resources are insufficient or channel noise is strong, transmitting the entire video stream can impair delivery and reconstruction quality.

Semantic communication (SemCom) has emerged as a promising paradigm for wireless networks by transmitting task-relevant information rather than pursuing exact bit-level recovery~\cite{getu2024semantic,peng2026simac}. Recent studies have extended SemCom to video transmission by exploiting semantic importance, temporal redundancy, and adaptive transmission. Bao et al.~\cite{bao2025mdvsc} proposed MDVSC, which employs semantic importance coding for efficient wireless video transmission. Zhang et al.~\cite{zhang2026motion} designed a bidirectional motion-enhanced method to exploit temporal information in wireless video transmission. Yin et al.~\cite{yin2026generative} used multimodal semantic fusion for generative video transmission. Li et al.~\cite{li2026goal} developed a goal-oriented approach based on generative artificial intelligence. Ding et al.~\cite{ding2026foundation} developed a foundation model-aided channel-adaptive video SemCom system and evaluated it through a prototype. Peng et al.~\cite{peng2026sccvs} integrated computer vision sensing with SemCom to adapt edge video transmission to scene changes. These studies demonstrate the potential of content-aware and adaptive SemCom for efficient wireless video transmission.

These advances motivate two design challenges for importance-aware token transmission under bandwidth constraints:
\begin{enumerate}
\item \textbf{Modeling video content importance at the token level.}
Different regions and tokens in a video contribute unequally to the information relevant to an application. For example, moving objects and changing regions may require more frequent updates than slowly changing backgrounds. Without translating these differences into token priorities, limited communication resources may be allocated to content that requires fewer updates.
\item \textbf{Adapting token transmission to available bandwidth.}
Transmitting a fixed set of video tokens may exceed the available capacity as bandwidth changes. For example, a decrease in bandwidth can prevent the link from delivering all tokens at the target frame rate. A transmission policy therefore needs to retain higher-priority content while reducing the transmission of lower-priority tokens.
\end{enumerate}

To address these challenges, we propose a token importance-aware semantic communication (Ti-SemCom) framework. Ti-SemCom models the importance of video regions and adapts token transmission to the available bandwidth. The main contributions are summarized as follows:

\begin{enumerate}
\item We develop a hierarchical importance modeling method for video transmission. Object and motion information is jointly extracted to identify communication-relevant regions in each frame. These regions are divided into three importance levels, which determine the priorities of the corresponding semantic tokens.

\item We design a bandwidth-adaptive token transmission strategy that selects content according to token importance and available bandwidth. Higher-importance token groups are retained first as communication resources become limited. The receiver fills untransmitted regions with content from a reconstructed reference frame, reducing transmission while retaining updates to higher-priority regions.

\item We conduct experiments on the CAVIAR dataset~\cite{saleem2026objtracknet} under different bandwidth and signal-to-noise ratio (SNR) conditions. The experiments jointly evaluate pixelwise and structural reconstruction quality, region-level performance, and transmission efficiency. At $5\%$ of the reference bandwidth, Ti-SemCom reduces transmitted data by $88.3\%$ relative to full-token transmission. Its reconstruction quality remains similar across the evaluated bandwidth ratios at each tested SNR.
\end{enumerate}

\section{System Model}

We consider a point-to-point wireless video transmission system in which a surveillance camera sends video to a monitoring station over a bandwidth-limited link. The video is represented as $\mathcal{V}=\{\mathbf{x}_i\}_{i=1}^{N_v}$, where $N_v$ is the total number of frames. Each frame $\mathbf{x}_i\in\mathbb{R}^{H\times W\times C}$ has height $H$, width $W$, and $C$ channels. The transmitter extracts semantic representations from each frame and selects tokens according to their importance and the available communication resources. The receiver decodes the received tokens and reconstructs the current frame, filling untransmitted regions with previously reconstructed visual content.

\subsection{Encoder}

For each input frame $\mathbf{x}_i$, the semantic encoder extracts a compact latent representation as
\begin{equation}
\mathbf{z}_i=f_{\theta}(\mathbf{x}_i),
\end{equation}
where $f_{\theta}(\cdot)$ denotes the semantic encoder parameterized by $\theta$. The encoded representation is divided into $N_T$ semantic tokens, forming the complete token set
\begin{equation}
\mathcal{T}_i=\{\mathbf{z}_{i,j}\}_{j=1}^{N_T},
\end{equation}
where $\mathbf{z}_{i,j}$ denotes the $j$-th semantic token of frame $i$.

Since video regions differ in their importance to the application, only a subset of tokens may be selected for transmission. Let $\mathcal{S}_i\subseteq\mathcal{T}_i$ denote the selected token set for frame $i$. Selection depends on token importance and the available bandwidth. Section~\ref{sec:framework} introduces the importance model and bandwidth-adaptive token selection mechanism in detail.

The selected semantic tokens are converted into a bit sequence and channel-encoded as
\begin{equation}
\mathbf{b}_i=\mathcal{C}(\mathcal{S}_i),
\end{equation}
where $\mathcal{C}(\cdot)$ represents the token-to-bit conversion and channel coding process. The channel-coded bits are subsequently modulated as
\begin{equation}
\mathbf{s}_i=\mathcal{M}(\mathbf{b}_i),
\end{equation}
where $\mathcal{M}(\cdot)$ denotes the modulation operation and $\mathbf{s}_i$ represents the transmitted signal.

\subsection{Wireless Channel}

The modulated signal is transmitted over a wireless channel. The received signal is modeled as
\begin{equation}
\mathbf{r}_i=\mathbf{h}\mathbf{s}_i+\mathbf{n}_i,
\end{equation}
where $\mathbf{h}$ denotes the channel coefficient and $\mathbf{n}_i$ denotes additive noise. This model describes the signal received after wireless transmission. The numerical experiments use an additive white Gaussian noise (AWGN) channel.

\subsection{Decoder}

At the receiver, the received signal is first demodulated and channel-decoded as
\begin{equation}
\mathbf{y}_i=\mathcal{C}^{-1}\big(\mathcal{M}^{-1}(\mathbf{r}_i)\big),
\end{equation}
where $\mathcal{M}^{-1}(\cdot)$ denotes demodulation, while $\mathcal{C}^{-1}(\cdot)$ includes channel decoding and semantic token recovery. The recovered representation $\mathbf{y}_i$ is fed into the semantic decoder to obtain the current-frame reconstruction:
\begin{equation}
\tilde{\mathbf{x}}_i=g_{\phi}(\mathbf{y}_i),
\end{equation}
where $g_{\phi}(\cdot)$ denotes the semantic decoder parameterized by $\phi$.

Since only a subset of semantic tokens may be transmitted, some regions in $\tilde{\mathbf{x}}_i$ lack newly transmitted information. The receiver fills these regions with content from a reconstructed reference frame to obtain the final reconstruction $\hat{\mathbf{x}}_i$. Section~\ref{sec:framework} describes this reuse operation after semantic decoding.

\subsection{Problem Formulation}

The objective is to minimize average video reconstruction distortion subject to the available transmission capacity. For each frame, the transmitted subset $\mathcal{S}_i$ is selected from the complete token set $\mathcal{T}_i$. The optimization problem is formulated as
\begin{equation}
\min_{\{\mathcal{S}_i\subseteq\mathcal{T}_i\}}
\frac{1}{N_v}\sum_{i=1}^{N_v}D(\mathbf{x}_i,\hat{\mathbf{x}}_i),
\end{equation}
subject to
\begin{equation}
\sum_{i=1}^{N_v}|\mathcal{S}_i|S_T\leq B\frac{N_v}{F},
\end{equation}
where $D(\cdot)$ denotes distortion between the original and reconstructed frames, and $|\mathcal{S}_i|$ is the number of transmitted tokens for frame $i$. The transmission cost per token is $S_T$ bits, and $F$ denotes the target frame rate. Throughout this paper, bandwidth $B$ refers to the available transmission rate in bits per second. The constraint limits the total token transmission cost to the capacity available over the video duration $N_v/F$.

\begin{figure*}[htbp]
    \centering
    \includegraphics[width=17cm]{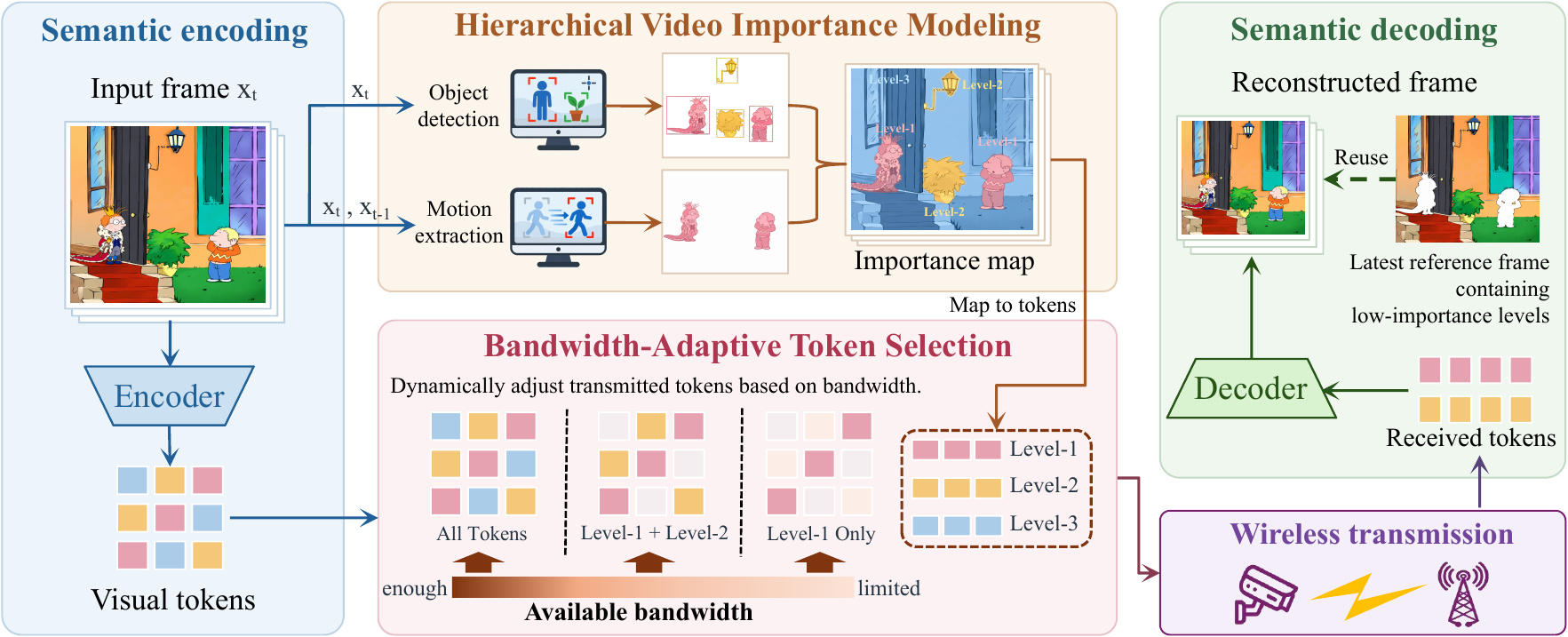}
    \caption{Overall architecture of the proposed Ti-SemCom framework.}
    \label{fig}
\end{figure*}

Under this formulation, transmitting more semantic tokens generally provides more information for reconstruction but incurs a higher communication cost. To address this problem, Ti-SemCom assigns hierarchical importance levels to tokens and adjusts the selected token groups according to available bandwidth.

\section{The Proposed Ti-SemCom Framework}
\label{sec:framework}

As illustrated in Fig.~\ref{fig}, Ti-SemCom jointly uses object and motion information to divide video regions into three importance levels. These levels are spatially associated with the semantic tokens generated by the encoder. Higher-importance token groups are selected first, subject to the available bandwidth. At the receiver, transmitted tokens are decoded to reconstruct the current frame. Untransmitted regions are filled with content at the same spatial positions in a reference frame, reducing the need to transmit lower-priority regions.

\subsection{Hierarchical Video Importance Modeling}

Different video regions contribute unequally to the information relevant to an application. Motion regions reflect changes between consecutive frames, while detected objects identify content that may be relevant to the monitoring task. We therefore combine object and motion information to construct a hierarchical importance representation.

Given the current frame $\mathbf{x}_t$ and the previous frame $\mathbf{x}_{t-1}$, the object and motion regions are obtained as
\begin{equation}
M_o=f_o(\mathbf{x}_t), \qquad M_m=f_m(\mathbf{x}_t,\mathbf{x}_{t-1}),
\end{equation}
where $M_o$ and $M_m$ are binary masks of the detected object and motion regions, respectively. Based on application relevance and temporal variation, the video regions are divided into three importance levels.

\textit{Level-1 Importance Region:}
Level-1 includes task-relevant critical objects and detected motion regions and receives the highest transmission priority:
\begin{equation}
M_1=M_{\mathrm{critical}}\lor M_m,
\end{equation}
where $M_{\mathrm{critical}}\subseteq M_o$ denotes critical object regions specified for the application, and $\lor$ denotes logical OR. These regions receive the highest priority because of their application relevance or temporal changes.

\textit{Level-2 Importance Region:}
Level-2 contains the remaining detected object regions that are not assigned to Level-1:
\begin{equation}
M_2=M_o\land\neg M_1.
\end{equation}

These regions contain detected objects and receive higher priority than Level-3 regions. Here, $\land$ and $\neg$ denote logical AND and NOT.

\textit{Level-3 Importance Region:}
Level-3 contains the remaining regions outside the detected object and motion masks:
\begin{equation}
M_3=\neg(M_1\lor M_2).
\end{equation}

These regions are not identified as objects or motion and receive the lowest transmission priority in the proposed hierarchy.

Based on this hierarchy, the semantic tokens associated with each region are assigned corresponding importance levels for bandwidth-adaptive token selection.

\subsection{Bandwidth-Adaptive Token Selection}

Building on the hierarchical importance map, we adapt token transmission to the time-varying available bandwidth. This policy gives priority to higher-importance video regions when communication resources are constrained.

The hierarchical importance map is spatially aligned with the semantic token grid before token-level assignment. Each token $\mathbf{z}_{i,j}$ is assigned an importance level $c_{i,j}$ based on its corresponding spatial region. This mapping converts region-level importance into token-level transmission priority. The tokens are grouped as
\begin{equation}
\mathcal{T}_{k,i}=\{\mathbf{z}_{i,j}\in\mathcal{T}_i\mid c_{i,j}=k\},\quad k\in\{1,2,3\},
\end{equation}
where $\mathcal{T}_{1,i}$, $\mathcal{T}_{2,i}$, and $\mathcal{T}_{3,i}$ contain Level-1, Level-2, and Level-3 tokens, respectively.

The token set is adjusted according to the bandwidth $B_i$ available for frame $i$. Let $D_{1,i}$, $D_{12,i}$, and $D_{123,i}$ denote the requirements for transmitting Level-1, Levels 1--2, and all three token levels, respectively. The requirements use the same units as $B_i$. The transmission state is determined as
\begin{equation}
q_i=
\begin{cases}
3, & B_i\ge D_{123,i},\\
2, & D_{12,i}\le B_i<D_{123,i},\\
1, & D_{1,i}\le B_i<D_{12,i},\\
0, & B_i<D_{1,i}.
\end{cases}
\end{equation}
For $q_i>0$, $q_i$ is the largest retained level index; $q_i=0$ indicates that no tokens are transmitted. For a positive transmission state, the transmitted token set is given by
\begin{equation}
\mathcal{S}_i=\bigcup_{k=1}^{q_i}\mathcal{T}_{k,i}\subseteq\mathcal{T}_i.
\end{equation}

As bandwidth decreases, lower-importance token groups are excluded according to the thresholds above. Selection operates at the importance-level granularity, with all tokens in each retained level transmitted together. Tokens are not ranked within a level, and $\mathcal{S}_i=\varnothing$ when $q_i=0$.

\subsection{Receiver-Side Reference Reuse}
% Author check: specify reference initialization and refresh conditions, and how the receiver obtains the current-frame mask.
For regions whose tokens are not transmitted in the current frame, the receiver reuses content from a reference frame. The reference is the most recent frame reconstructed from all three token levels. In particular, Level-3 tokens are transmitted only when all levels are selected, and omitted Level-3 regions are filled from this reference. Reuse is performed in the RGB domain after semantic decoding. According to the current-frame importance map, transmitted regions use the current decoder output $\tilde{\mathbf{x}}_i$. Untransmitted regions use content at the same spatial positions in the reference frame. Combining these regions produces the final reconstructed frame $\hat{\mathbf{x}}_i$.

\begin{figure*}[htbp]
\centering
% Keep the third panel's y-axis labels clear of the intercolumn gutter.
\makebox[\textwidth][r]{\includegraphics[width=0.98\textwidth]{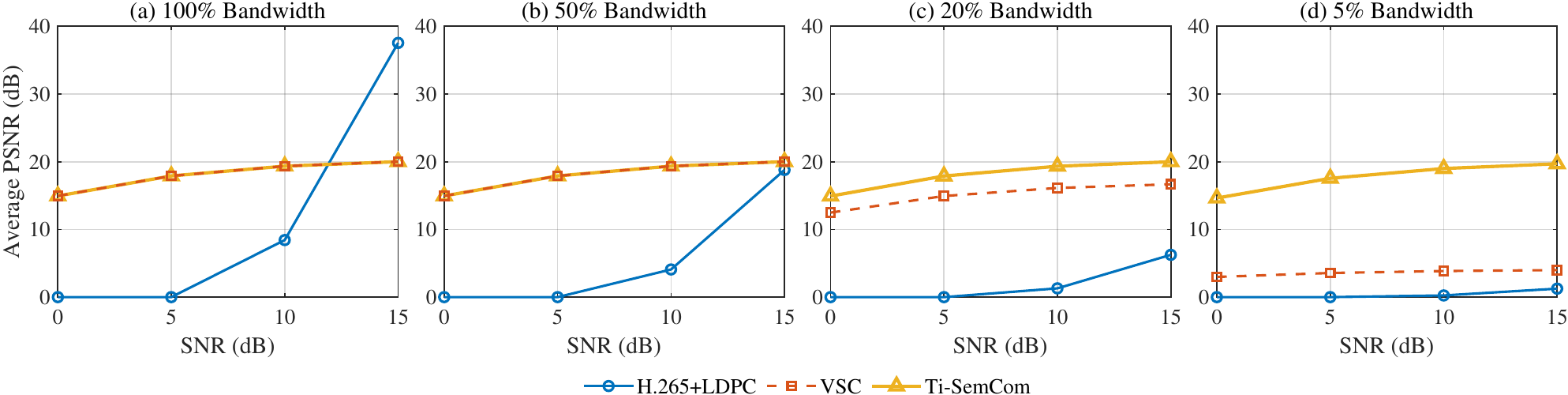}}
\caption{PSNR comparison results under different SNR and bandwidth conditions. (a) 100\% bandwidth. (b) 50\% bandwidth. (c) 20\% bandwidth. (d) 5\% bandwidth.}
\label{fig:snr}
\end{figure*}

\begin{figure*}[htbp]
\centering
% Keep the third panel's y-axis labels clear of the intercolumn gutter.
\makebox[\textwidth][r]{\includegraphics[width=0.98\textwidth]{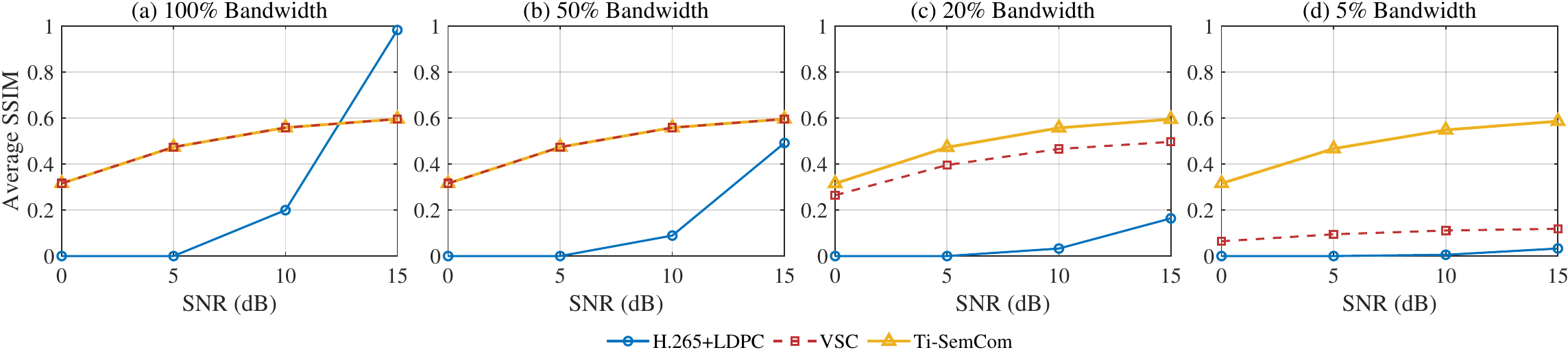}}
\caption{Average SSIM of H.265+LDPC, VSC, and Ti-SemCom under different SNR and bandwidth conditions. (a) 100\% bandwidth. (b) 50\% bandwidth. (c) 20\% bandwidth. (d) 5\% bandwidth.}
\label{fig:ssim}
\end{figure*}

\begin{figure}[htbp]
\centering
\includegraphics[width=0.95\columnwidth]{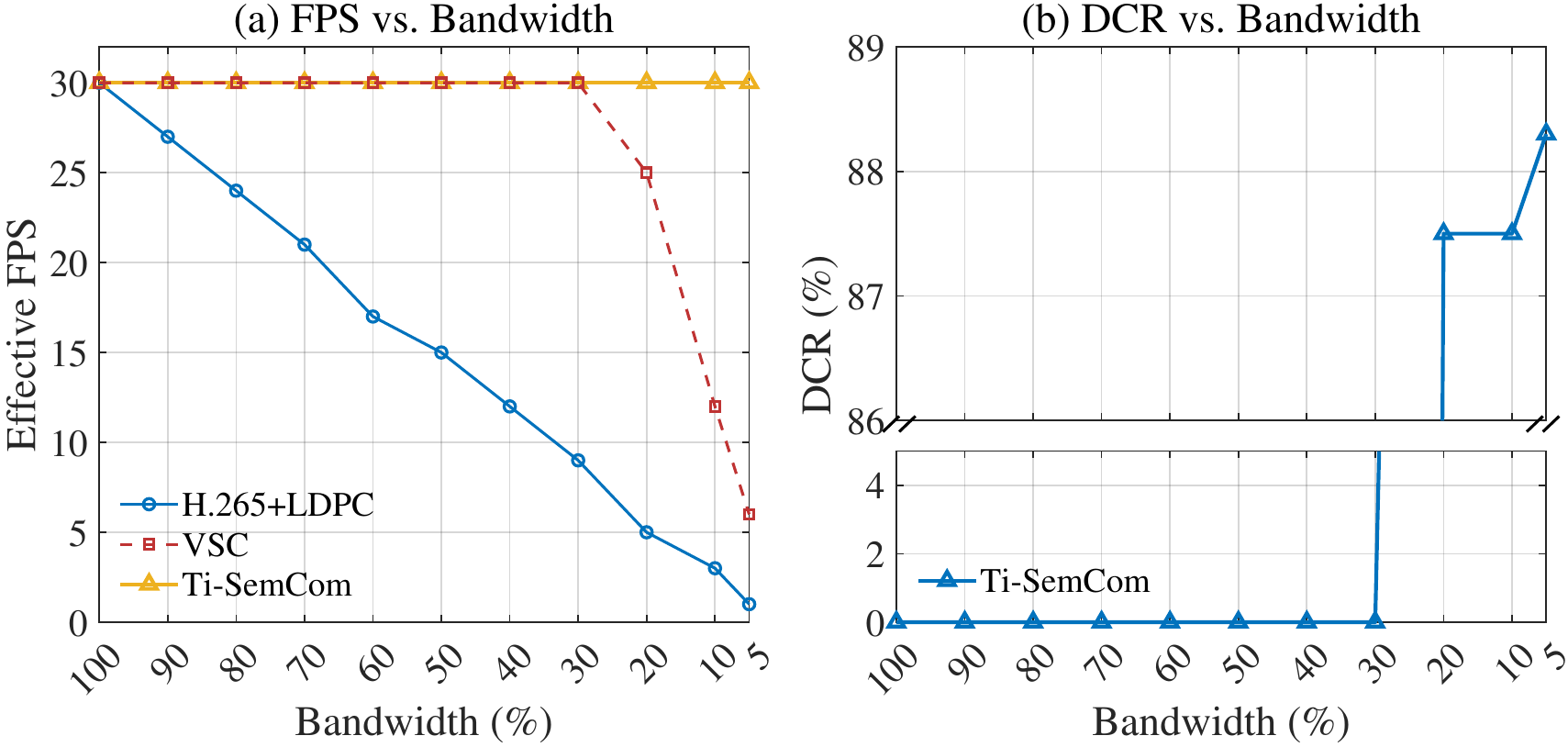}
\caption{Bandwidth-aware transmission performance of H.265+LDPC, VSC, and Ti-SemCom. (a) Effective FPS. (b) Data Compression Ratio.}
\label{fig:bandwidth}
\end{figure}

\subsection{Model Training}
% Author check: specify mask construction and whether the loss is evaluated before or after RGB reference reuse.
During training, predefined importance masks expose the semantic model to different token retention states. During inference, the hierarchical importance map is dynamically generated from object and motion information. To emphasize reconstruction in higher-importance regions, we use an importance-aware weighted loss (IWL). The reconstruction loss for each importance level is the mean squared error (MSE):
\begin{equation}
L_k=\mathrm{MSE}(x^{(k)},\hat{x}^{(k)}), \quad k\in\{1,2,3\},
\end{equation}
where $x^{(k)}$ and $\hat{x}^{(k)}$ denote the original and reconstructed pixel values in the Level-$k$ region, respectively. The overall training loss is
\begin{equation}
L=\sum_{k=1}^{3}w_kL_k,
\end{equation}
where $w_k$ denotes the weight assigned to the Level-$k$ reconstruction loss. Larger relative weights can be assigned to higher-importance regions to give their reconstruction errors greater influence on the training objective.

\section{Experimental Results}

\subsection{Experimental Setup}
We use the CAVIAR dataset for evaluation~\cite{saleem2026objtracknet}. It contains video sequences recorded in indoor public environments at a resolution of $384\times288$ pixels. The source frame rate is 25 frames per second (FPS).

The semantic encoder converts each frame into 256 semantic tokens with a feature dimension of 192. YOLOv8-Seg~\cite{10967648} and MOG2~\cite{10460003} extract object and motion masks, respectively, and the importance map is resized to $28\times28$. The target transmission frame rate is 30 FPS. The H.265 baseline uses low-density parity-check (LDPC) channel coding. The reference bandwidth is $B_{\mathrm{ref}}=9.77$ Mbps, the minimum rate required for H.265+LDPC to maintain 30 FPS in the evaluation. Bandwidth percentages are relative to $B_{\mathrm{ref}}$. All methods use an AWGN channel with quadrature phase-shift keying (QPSK) modulation. The SNR experiment uses values of 0, 5, 10, and 15 dB.

\subsection{Compared Methods}
We compare the following video communication schemes under different bandwidth and SNR conditions:

\begin{itemize}
\item \textbf{H.265+LDPC:} High Efficiency Video Coding (HEVC/H.265)~\cite{11134514} with LDPC channel coding~\cite{10879552}.

    \item \textbf{Video SemCom (VSC):} The baseline SemCom system for video transmission, in which all semantic tokens are transmitted without importance-based selection.

    \item \textbf{Ti-SemCom:} The proposed framework for importance-aware video transmission.
\end{itemize}

\subsection{Evaluation Metrics}
% Author check: define the PSNR/SSIM treatment of untransmitted or failed frames and the criterion for successful frame transmission.

We evaluate reconstruction quality and bandwidth efficiency using the following metrics.

\begin{itemize}
\item \textbf{Reconstruction Quality:}
For each evaluated frame, we compute the peak signal-to-noise ratio (PSNR) and structural similarity (SSIM) between the original and reconstructed frames. Average PSNR and SSIM are obtained by taking the arithmetic mean of the corresponding per-frame values over all evaluated frames.

\item \textbf{Bandwidth Efficiency:}
Effective FPS measures the number of successfully transmitted video frames per second:
\begin{equation}
\mathrm{Effective\ FPS}=\frac{N_{\mathrm{true}}}{T},
\end{equation}
where $N_{\mathrm{true}}$ counts successfully transmitted frames during an interval of $T$ seconds. For Ti-SemCom, the data compression ratio (DCR) measures the data reduction achieved by omitting lower-importance tokens:
\begin{equation}
\mathrm{DCR}=\left(
1-\frac{D_{\mathrm{tx}}}{D_{\mathrm{full}}}
\right)\times100\%,
\end{equation}
where $D_{\mathrm{full}}$ and $D_{\mathrm{tx}}$ denote full-token and actual transmitted data volumes over the same video interval, respectively. A higher DCR indicates greater transmitted-data reduction relative to full-token transmission.
\end{itemize}

\subsection{Simulation Evaluation}

\begin{figure*}[htbp]
\centering
\includegraphics[width=17cm]{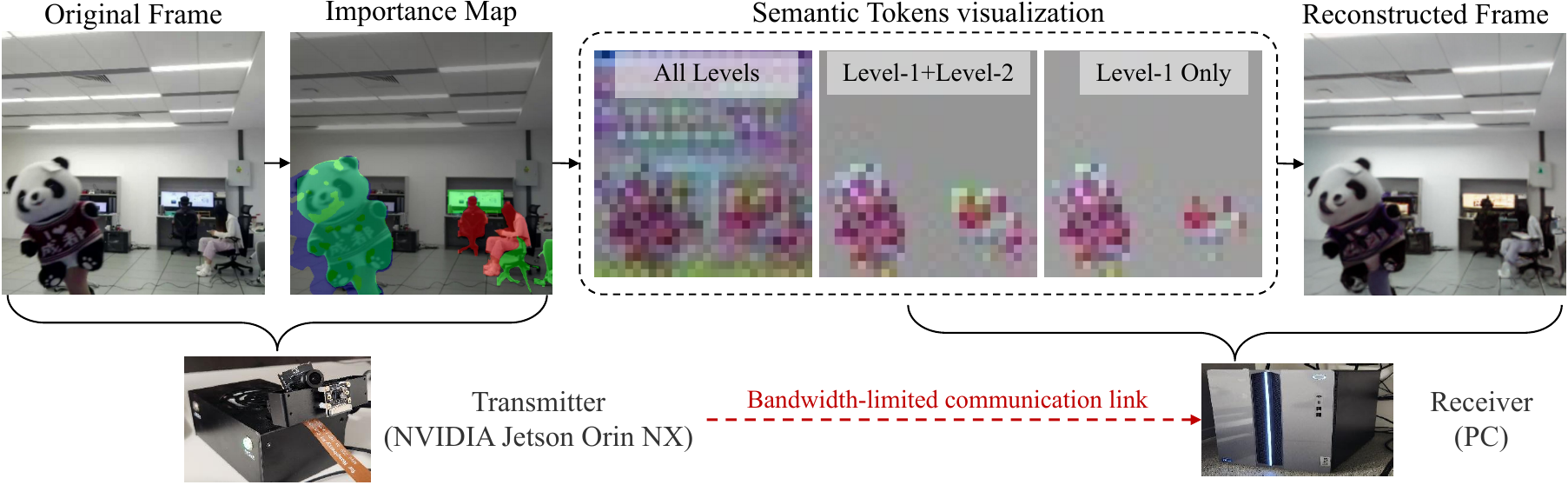}
\caption{Prototype evaluation of Ti-SemCom under low bandwidth conditions.}
\label{fig:qualitative}
\end{figure*}

The proposed Ti-SemCom framework is evaluated in terms of SNR, bandwidth, and region-level reconstruction performance. PSNR and SSIM provide complementary comparisons of pixelwise and structural reconstruction quality, respectively.

\subsubsection{Reconstruction Quality vs. SNR}
Figs.~\ref{fig:snr} and~\ref{fig:ssim} compare reconstruction quality in terms of PSNR and SSIM under different SNR and bandwidth conditions. H.265+LDPC achieves high reconstruction quality at high SNR with sufficient bandwidth, but both H.265+LDPC and VSC degrade as bandwidth decreases. In contrast, Ti-SemCom maintains similar PSNR and SSIM across the tested bandwidth ratios at each SNR and outperforms both baselines at $20\%$ and $5\%$ bandwidth. These results are consistent with its importance-based selection policy, which prioritizes higher-importance tokens when bandwidth is limited, helping preserve both pixelwise and structural reconstruction quality.

\subsubsection{Transmission Efficiency vs. Bandwidth}
As shown in Fig.~\ref{fig:bandwidth}, reducing bandwidth from $100\%$ to $5\%$ decreases the effective frame rates of H.265+LDPC and VSC from 30 FPS to 1 and 6 FPS, respectively. In contrast, Ti-SemCom maintains 30 FPS across the tested bandwidth conditions, with a DCR of $87.5\%$ at $20\%$ and $10\%$ bandwidth and $88.3\%$ at $5\%$ bandwidth relative to full-token transmission. By selectively omitting lower-importance token groups, Ti-SemCom reduces transmitted data while retaining higher-priority content and maintaining the target effective frame rate.

\subsubsection{Reconstruction Quality vs. Importance Weights}
We evaluate IWL without token dropping, with all semantic tokens retained. Table~\ref{tab:iwl} reports region-level and full-frame reconstruction quality under different importance weight ratios.

\begin{table}[htbp]
\centering
\caption{PSNR (dB) under different importance weights.}
\label{tab:iwl}
\resizebox{\columnwidth}{!}{
\begin{tabular}{c|cccc}
\hline
$w_1:w_2:w_3$ & Level-1 & Level-2 & Level-3 & Full frame \\
\hline
$1:1:1$   & 17.12 & 23.96 & 21.05 & 20.94 \\
$2:1:1$   & 17.40 & 22.45 & 19.51 & 19.46 \\
$3:1.5:1$ & 17.58 & 22.37 & 18.81 & 18.78 \\
$4:2:1$   & 17.66 & 22.33 & 18.39 & 18.36 \\
\hline
\end{tabular}}
\end{table}

Changing the weight ratio from $1:1:1$ to $4:2:1$ increases Level-1 PSNR from 17.12 to 17.66 dB, a gain of 0.54 dB. Full-frame PSNR decreases from 20.94 to 18.36 dB, a loss of 2.58 dB. These results show a trade-off between reconstruction quality in high-priority regions and overall frame quality.

\subsection{Prototype Evaluation}
We implement a video transmission prototype to evaluate Ti-SemCom on an edge transmitter and a PC receiver. The transmitter is deployed on an NVIDIA Jetson Orin NX with 16 GB of memory. The receiver runs on a PC with an Intel Core i7-14700F CPU, an NVIDIA GeForce RTX 4060 Ti GPU, and 32 GB of RAM. The transmitter performs semantic encoding and importance-aware token selection, then sends the selected tokens through the bandwidth-limited link. The receiver decodes the tokens and applies reference-frame reuse to reconstruct the video. This setup evaluates operation of Ti-SemCom on the implemented edge-to-PC platform.

As shown in Fig.~\ref{fig:qualitative}, Ti-SemCom selects tokens according to the generated importance map. Higher-importance regions receive transmission priority when communication resources are limited. At a link rate of 288 kbps, the prototype achieves 8.4 FPS. This result demonstrates operation of the implemented platform under a bandwidth constraint; the figure provides a qualitative view of the reconstructed content.

\section{Conclusion}
This paper presents Ti-SemCom for robust video transmission under limited wireless resources. Object and motion information is jointly used to classify video regions into three importance levels. These levels determine the transmission priorities of the corresponding semantic tokens. Token selection adapts to the available bandwidth, while the receiver fills untransmitted regions with content from a reconstructed reference frame.
Experiments on the CAVIAR dataset show similar reconstruction quality, measured by PSNR and SSIM, across the evaluated bandwidth ratios at each tested SNR. At $5\%$ bandwidth, Ti-SemCom reduces transmitted data by $88.3\%$ relative to full-token transmission while maintaining the target effective frame rate. Future work will investigate task-driven adaptive importance modeling to determine token priorities for different application requirements.

\bibliographystyle{IEEEtran}
\bibliography{references}

\end{document}